\documentclass[ aip, jmp, amsmath,amssymb, reprint, ]{revtex4-1}

\usepackage{amssymb}
\usepackage{amsfonts}
\usepackage{amsmath}
\usepackage{amsthm}
\usepackage{epsfig}
\usepackage{color} 
\usepackage{subfigure}

\graphicspath{{./}{./Figures/}}

\begin{document}

\title{Chimera states and the interplay between initial conditions and non-local coupling}

\author{Peter Kalle}
\affiliation{Institut f{\"u}r Theoretische Physik, Technische Universit{\"a}t Berlin, Hardenbergstra\ss{}e 36, 10623 Berlin, Germany}
\author{Jakub Sawicki}
\affiliation{Institut f{\"u}r Theoretische Physik, Technische Universit{\"a}t Berlin, Hardenbergstra\ss{}e 36, 10623 Berlin, Germany}
\author{Anna Zakharova}
\affiliation{Institut f{\"u}r Theoretische Physik, Technische Universit{\"a}t Berlin, Hardenbergstra\ss{}e 36, 10623 Berlin, Germany}
\author{Eckehard Sch\"oll} 
\affiliation{Institut f{\"u}r Theoretische Physik, Technische Universit{\"a}t Berlin, Hardenbergstra\ss{}e 36, 10623 Berlin, Germany}

\date{\today}

\begin{abstract}

Chimera states are complex spatio-temporal patterns that consist of coexisting domains of coherent and incoherent dynamics. We study chimera states in a network of non-locally coupled Stuart-Landau oscillators. We investigate the impact of initial conditions in combination with non-local coupling. Based on an analytical argument, we show how the coupling phase and the coupling strength are linked to the occurrence of chimera states, flipped profiles of the mean phase velocity, and the transition from a phase- to an amplitude-mediated chimera state.

\end{abstract}


\maketitle

\begin{quotation}
Chimera states are an example of intriguing partial synchronization patterns appearing in networks of identical oscillators with symmetric coupling scheme. They exhibit a hybrid structure combining coexisting spatial domains of coherent (synchronized) and incoherent (desynchronized) dynamics, and were first reported for the model of phase oscillators\cite{KUR02a, ABR04}. Recent studies have demonstrated the emergence of chimera states in a variety of topologies, and for different types of individual dynamics\cite{PAN15,SCH16b}. In this paper, the interplay between initial conditions and non-local coupling is studied. We show that, based on an analytical argument incorporating the initial conditions and the range of non-local coupling, the occurrence of phase chimeras can be seen as caused by a phase lag in the coupling. Considering the dynamics of chimera states, our argument shows how ``flipped'' profiles of the mean phase velocities can be explained by a change of sign of the coupling phase. By this, one can either choose a concave (``upside'') profile of the mean phase velocities, or a ``flipped'' one. Extending our reasoning, we show that this argument intuitively explains the transition from phase- to amplitude-mediated chimera state as a result of increasing coupling strength.

\end{quotation}
\section{Introduction}

The analysis of coupled oscillatory systems is an important research field bridging between nonlinear dynamics, network science, and statistical physics, with a variety of applications in physics, biology, and technology\cite{PIK01,BOC06a}. The last decade has seen an increasing interest in chimera states in dynamical networks\cite{PAN15,SCH16b,LAI09}. First obtained in systems of phase oscillators\cite{KUR02a, ABR04}, chimeras can also be found in a large variety of different systems including time-discrete maps\cite{OME11,SEM15,VAD16}, time-continuous chaotic models\cite{OME12}, neural systems\cite{OME13,HIZ13,OME15,TSI16}, Boolean networks\cite{ROS14}, population dynamics\cite{HIZ15,BAN16}, Van der Pol oscillators\cite{OME15a, ULO16}, and quantum oscillator systems\cite{BAS15}. Moreover, chimera states allow for higher spatial dimensions\cite{OME12a,PAN15,SHI04,MAI15}. Together with the initially reported chimera states, which consist of one coherent and one incoherent domain, new types of these peculiar states having multiple\cite{OME13,VUE14,OME15a,SET08,XIE14} or alternating\cite{HAU15} incoherent regions, as well as amplitude-mediated\cite{SET13,SET14}, and pure amplitude chimera and chimera death states\cite{ZAK14,BAN15} were discovered. A classification has recently been given \cite{KEM16}.
In many systems, the form of the coupling defines the possibility to obtain chimera states. The nonlocal coupling has generally been assumed to be a necessary condition for chimera states to evolve in coupled systems. However, recent studies have shown that even global all-to-all coupling\cite{SET14,YEL14,BOE15,SCH15a,SCH15e}, as well as more complex coupling topologies allow for the existence of chimera states\cite{TSI16,HIZ15,KO08,OME15,ULO16}. Furthermore, time-varying network structures can give rise to alternating chimera states\cite{BUS15}. Chimera states have also been shown to be robust against inhomogeneities of the local dynamics and coupling topology\cite{OME15,LAI10}, as well as against noise\cite{LOO16}, or they might even be induced by noise\cite{SEM15b,SEM16}.
Possible applications of chimera states in natural and technological systems include the phenomenon of uni-hemispheric sleep\cite{RAT00,RAT16}, bump states in neural systems\cite{LAI01,SAK06a}, epileptic seizures\cite{ROT14}, power grids\cite{MOT13a}, or social systems\cite{GON14}. Many works considering chimera states have mostly been based on numerical results. A deeper bifurcation analysis\cite{OME13a} and even a possibility to control chimera states\cite{SIE14c,BIC15,OME16} were obtained only recently.
The experimental verification of chimera states was first demonstrated in optical\cite{HAG12} and chemical\cite{TIN12,NKO13} systems. Further experiments involved mechanical\cite{MAR13,KAP14}, electronic\cite{LAR13,GAM14}, optoelectronic delayed-feedback\cite{LAR15} and electrochemical\cite{WIC13,SCH14a} oscillator systems, Boolean networks\cite{ROS14}, and optical combs\cite{VIK14}.

Motivated by these studies, the goal of the present manuscript is to discuss how a specific set of initial
conditions initially separating the network into distinct domains gives rise to a clustered chimera state. This approach allows three statements to be validated: First, it will be discussed how this approach provides an intuitive
answer to the question why a pronounced off-diagonal coupling (a coupling
phase $\alpha$ close to $\pi/2$) is needed in order to access chimera
states. Second, it will be explained how a change of the sign of the
coupling phase $\alpha$ leads to the occurrence of normal and ``flipped''
arc-shaped profiles of the mean phase velocities, respectively.
These profiles are believed to be a distinct feature of (phase) chimeras, 
at least in the case of non-locally coupled systems. Third, it will 
be discussed how an increase of the coupling strength $\sigma$ is linked 
to a transition from a pure phase chimera state to a coupled phase-amplitude chimera 
state. The latter shows the main properties of an amplitude-mediated chimera 
state \cite{SET13}, i.e., the variations in the phases are connected with 
non-vanishing variations in the amplitudes.

\section{Model}

We consider a ring network of non-locally coupled Stuart-Landau oscillators. The local dynamics is given by the 
generic expansion (normal form) of an oscillator near a supercritical Hopf bifurcation

\begin{equation}
\dot{z}=(\lambda+i\omega)z-\left|z\right|^{2}z,
\end{equation}
where $\lambda \in \mathbb{R}$ is the bifurcation parameter, $\omega>0$ is the frequency
of the self-sustained oscillation and $z\in\mathbb{C}$ is the dynamical
variable. In the co-rotating frame\cite{GAR12b}, applying an appropriate scaling of time $t$, space $x$, and $z$,
\begin{eqnarray}
\tilde{t}&=&\lambda t,\label{eq:-0} \\
\tilde{x}&=&\lambda^{-1} x, \\
\tilde{z}&=&\lambda^{-1/2}\exp(-i\omega t) z,
\end{eqnarray}
and then dropping the tilde, the local dynamics is simplified to
\begin{equation}
\dot{z}=(1-\left|z\right|^{2})z=f(z),\label{eq:-1}
\end{equation}
where $\lambda>0$ has been assumed.
The network can be described in the continuum limit by the following partial differential equation,

\begin{equation}
\partial_{t}z(x,t)=f(z)+\sigma e^{i\alpha}\int_{0}^{L}G(x-x^{\prime})\left[z(x^{\prime})-z(x)\right]dx^{\prime},\label{eq:3}
\end{equation}
where the local dynamics $f(z)$ of an oscillator is given by Eq.\,(\ref{eq:-1}), $\sigma$
is the coupling strength, $\alpha$ is the coupling phase, $L$ is the system size assuming periodic boundary conditions, and $G(x-x^{\prime})$ is the coupling kernel determining the functional shape and range
of the non-local coupling. Here we assume that the kernel is given by a Gaussian with mean zero 

\begin{equation}
\label{eq:4}
G(x-x^{\prime})=c\, e^{-\left|x-x^{\prime}\right|^{2}},
\end{equation}
where $c=1/\Gamma(\tfrac{1}{2})$ denotes the normalization factor and $\Gamma$ is the gamma-function,
but our results hold also for more general kernels.

To motivate a specific choice of parameters and initial conditions
governing the emergence of chimera states, the system is transformed to polar coordinates via
$z=r\,\exp(i\theta)$. This yields the following partial differential
equations that describe the evolution of the amplitude $r$ and phase $\theta$,
\begin{widetext}
\begin{eqnarray}
\partial_{t}r(x,t) & = & F(r)+\underbrace{\sigma\int_{0}^{L}G(x-x^{\prime})\left[r(x^{\prime})\cos(\theta(x^{\prime})-\theta(x)+\alpha)-r(x)\cos(\alpha)\right]dx^{\prime}}_{\Sigma_{r}},\label{eq:5}\\
\partial_{t}\theta(x,t) & = & \underbrace{\sigma\int_{0}^{L}G(x-x^{\prime})\left[\frac{r(x^{\prime})}{r(x)}\sin(\theta(x^{\prime})-\theta(x)+\alpha)-\sin(\alpha)\right]dx^{\prime}}_{\Sigma_{\theta}}.\label{eq:6}
\end{eqnarray}
\end{widetext}
The local dynamics of the amplitudes is given by $F(r)=(1-r^{2})r$ 
with a stable fixed point $r_{0}=1$. 
In the following we study the impact
of the non-local coupling on the dynamics of the network. Introducing the amplitude coupling $\Sigma_{r}$ and the phase coupling 
$\Sigma_{\theta}$ we can write Eqs.\,(\ref{eq:5}) and (\ref{eq:6}) as 
\begin{eqnarray}
\partial_{t}r(x,t) & = & F(r)+\Sigma_{r}(x,t),\label{eq:7}\\
\partial_{t}\theta(x,t) & = & \Sigma_{\theta}(x,t).\label{eq:8}
\end{eqnarray}
\\

For the numerical simulations we use the discretized version of Eq.\,(\ref{eq:3}), i.e., a ring of $N$ coupled oscillators
\begin{eqnarray}
\label{eq:1}
\dot{z}_{j}&=&f(z_{j})+\sigma e^{i\alpha}\sum_{k=1}^{N}G_{jk}\left[z_{k}-z_{j}\right],
\end{eqnarray}
where $j=1,...,N$ and all indices are modulo $N$. $G_{jk}=\Delta x \, G\left(\Delta x\, [j-k]\right)$ is the discretized version of the coupling kernel in Eq.\,(\ref{eq:4}), where $\Delta x = L/N$ is the spatial increment between neighboring oscillators.

\section{The impact of initial conditions}

An important issue, often considered as a necessary condition for
the existence of chimera states, is the choice of initial conditions.
Random initial conditions do not always guarantee chimera behavior.
This is due to the fact that classical chimera states typically coexist
with the completely synchronized regime. In the case of chimera states
the basin of attraction can be relatively small in comparison with
that of the synchronized state. In the present work we discuss the impact of specially prepared initial conditions and non-local coupling in order to explain, predict and confirm the occurrence of chimera states and their main features. 

\subsection{From initial conditions to a clustered chimera state}
\begin{figure}
\begin{centering}
\includegraphics[width=0.4\textwidth]{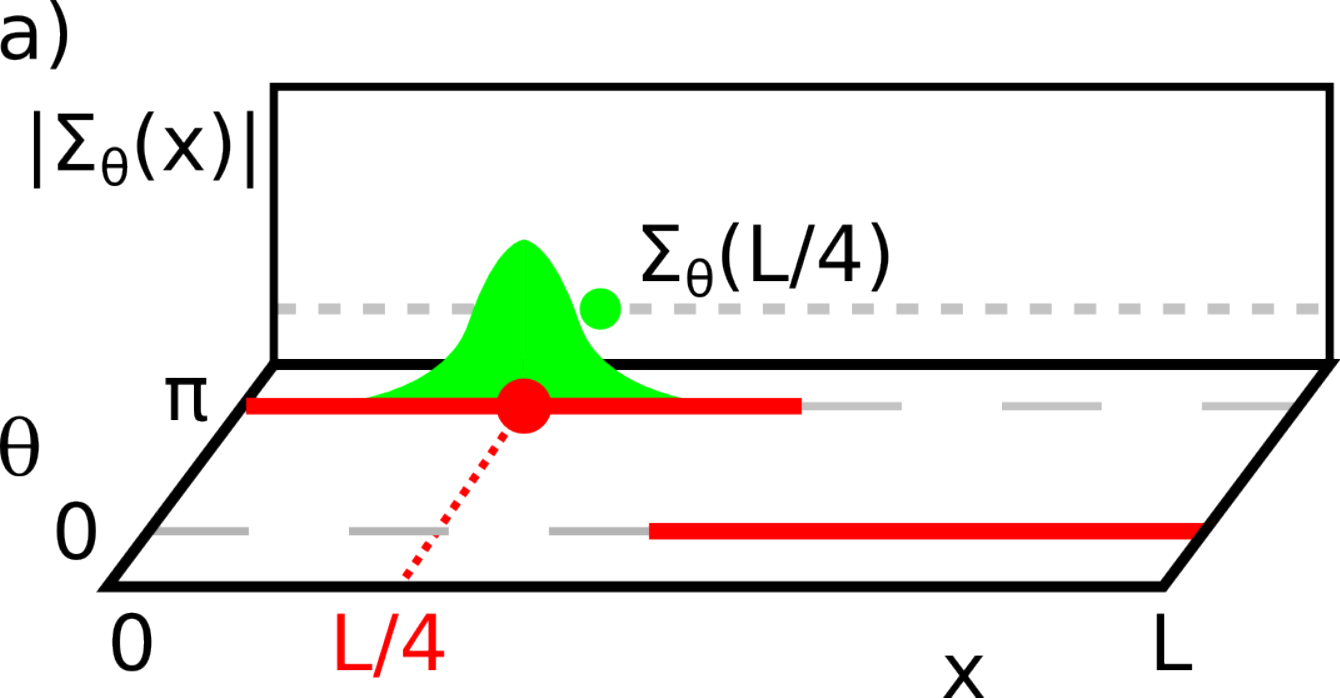}
\par\end{centering}

\begin{centering}
\includegraphics[width=0.4\textwidth]{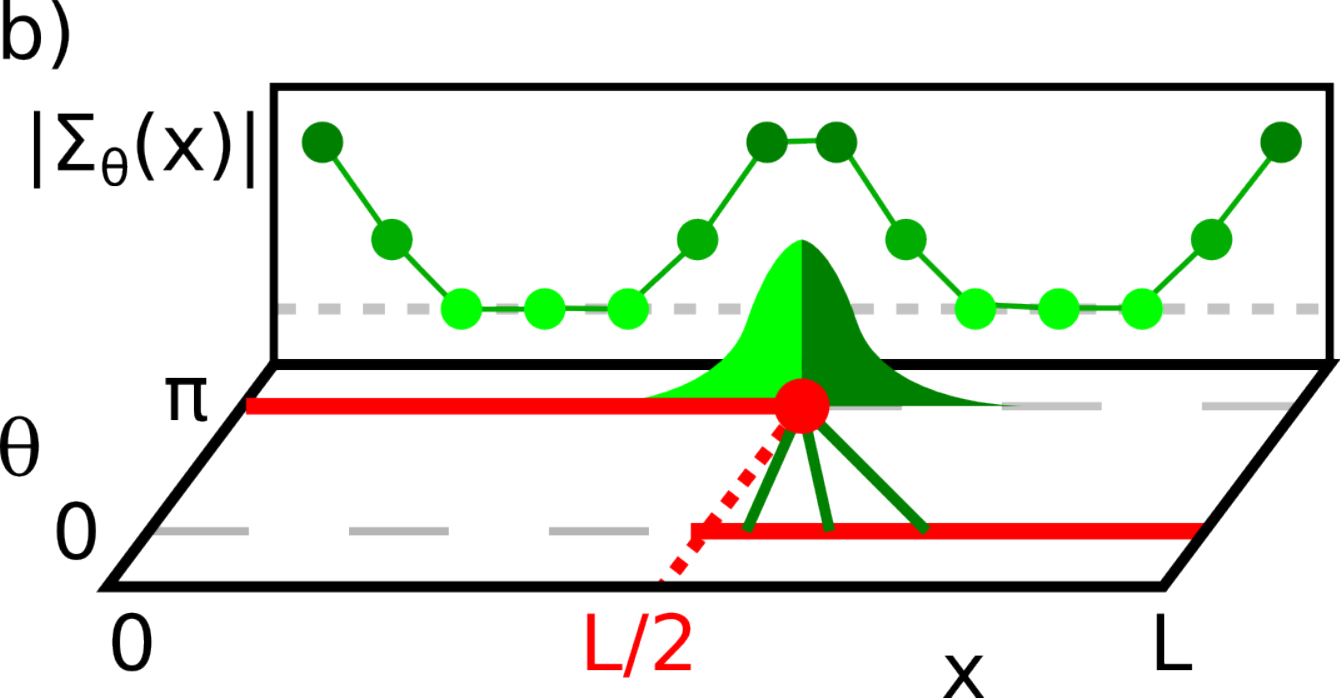}
\par\end{centering}

\begin{centering}
\includegraphics[width=0.4\textwidth]{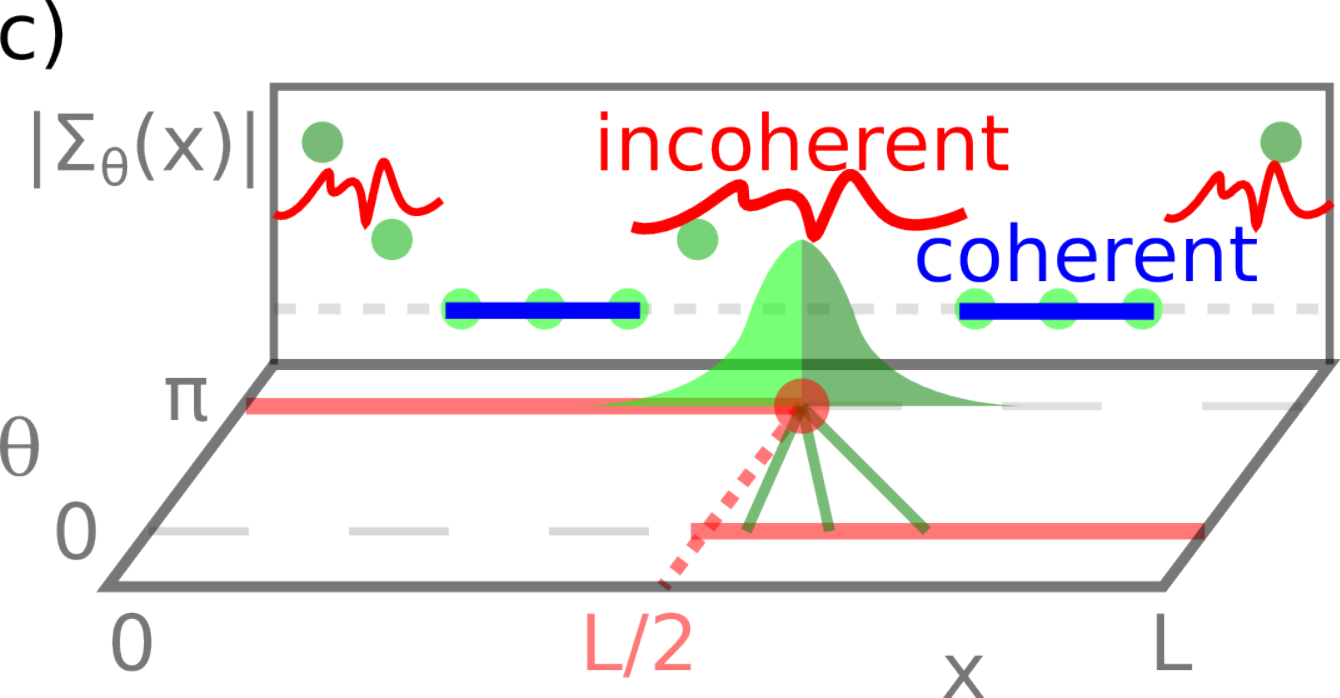}
\par\end{centering}

\caption{Sketch of the initial dynamical scenario obtained by the choice
of two populations initially in anti-phase, as given by Eq.\,(\ref{eq:91}).
The distribution of the phases $\theta$ vs space $x$ is shown by full red lines. The coupling kernel $G(x-x^{\prime})$ localized at a specific oscillator (red dot) is shaded (green).  
(a) Oscillator at the center of an in-phase population at $x_{0}=L/4$, yielding a vanishing coupling term $\Sigma_\theta=0$. 
(b) Oscillator at the border between in-phase and anti-phase populations $x_{c}=L/2$, yielding a maximum coupling term 
$\Sigma_\theta(x_c)$  The green connected dots sketch the profile of the coupling term $\Sigma_\theta$ vs $x$. The magnitude of the initial coupling term $\Sigma_\theta(x)$ is illustrated by the brightness of the green color. There are four distinct regions, two where the coupling term nearly vanishes, and two where it does not (note the periodic boundary conditions in $x$). 
(c) Sketch of the dynamical scenario arising from this distribution of the initial coupling term. The blue straight lines illustrate coherent states with a constant phase, where the phase dynamics of the oscillators is not perturbed by the coupling term. The red twisted lines denote incoherent states with varying phases. These are centered around the borders between the two oscillator population. In these regions the coupling term does not vanish due to the non-local coupling to oscillators in anti-phase. \label{fig:1}}
\end{figure}

\begin{figure}
\begin{centering}
\includegraphics[width=0.5\textwidth]{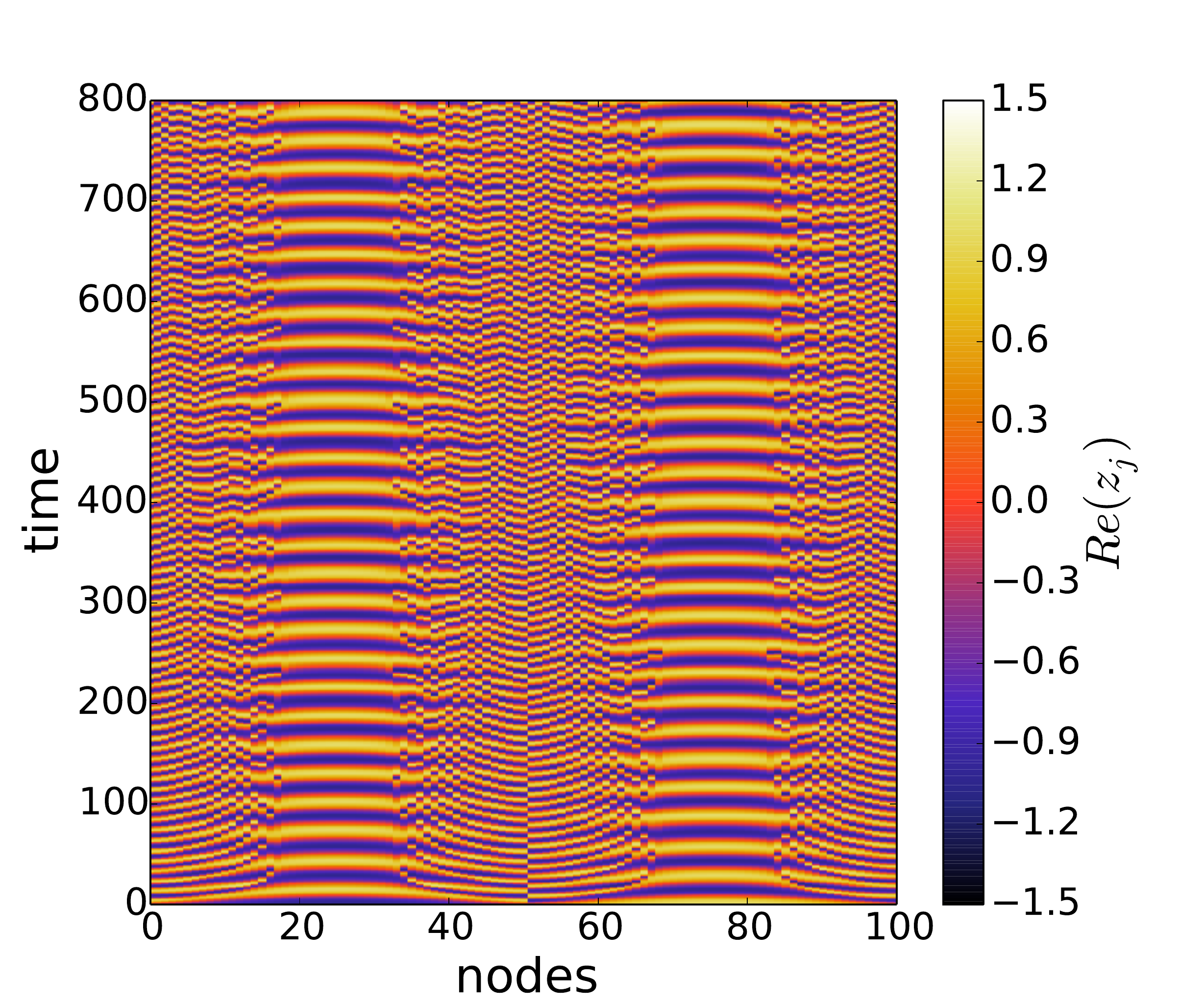}
\par\end{centering}

\centering{}\caption{Space-time plot of $Re(z_{j})$ in a network of $N$ non-locally coupled oscillators. The initial conditions are given by Eqs.\,(\ref{eq:9}) and (\ref{eq:91}). Parameters: $\sigma=0.6$, $\alpha=\pi/2-0.15$, $N=101$, $L=2\pi$.
\label{fig:1-1-3}}
\end{figure}

Using an anti-phase cluster as initial condition, it is possible
to simplify the initial coupling terms in amplitude and phase significantly.
The initial conditions are chosen as two clusters in anti-phase,

\begin{eqnarray}
r(x,t_{0}) & = & 1,\label{eq:9}\\
\theta(x,t_{0}) & = & \begin{cases}
\pi & ,\, if\, x\in(0,L/2]\\
0 & ,\, if\, x\in(L/2,L]
\end{cases}.\label{eq:91}
\end{eqnarray}
The network is initially divided into two equally sized domains. The
first one, with phase $\pi$, reaches from $0$ to $L/2$. The second
one, with phase $0$, reaches from $L/2$ to $L$. By this choice
of two domains in anti-phase, the network is initially spatially separated 
into four distinct domains with respect to the coupling terms $\Sigma_{r}$
and $\Sigma_{\theta}$. This is schematically shown in Fig.\,\ref{fig:1}.
Two domains, where the coupling terms vanish because the oscillators
are coupled solely to oscillators in phase (Fig.\,\ref{fig:1}a),
are separated by two domains where the coupling terms
$\Sigma_{r}$ and $\Sigma_{\theta}$ have finite, non-vanishing
values due to the coupling to oscillators that are in anti-phase (Fig.\,\ref{fig:1}b). This initial separation influences the corresponding
long-time behavior significantly. While the dynamics of the two populations
with almost vanishing coupling terms becomes synchronized, the two populations
where the coupling does not vanish initially, are perturbed in their
phase and amplitude dynamics, see Fig.\,\ref{fig:1}c. The corresponding
chimera state can be clearly seen in a space-time plot, where the
dynamics is shown for the real parts $Re(z_{j})$ for every
node of the network (Fig.\,\ref{fig:1-1-3}). The two populations
of oscillators being initially in anti-phase split into the four domains
mentioned. Two clusters in anti-phase are formed around the centers of the
initial in-phase domains at $x=L/4$ $(j=25)$ and $x=3L/4$ $(j=75)$. The two coherent 
domains are separated by incoherent domains, their initial centers being at
$x=L/2$ $(j=50)$ and  $x=L$ $(j=100)$. 
 
The validity of this approach has been tested for long simulation times 
and increasing numbers of oscillators forming the network. Our simulations confirm that the observed chimera states are long-living and rule out finite-size effects for oscillator numbers up to $N=1001$.

\subsection{Off-diagonal coupling revisited}

It has been shown recently that a value of the coupling phase $\alpha\simeq\pi/2$
is required for chimera states in phase oscillators \cite{ABR04,OME13}. Also the chimera
state shown in Fig.\,\ref{fig:1-1-3} requires a coupling phase $\alpha$
close to $\pi/2$ in order to be observed. Such a condition has been used in many studies \cite{ABR04,OME13,SIE14c}.
We show that the approach outlined in the previous section gives an intuitive explanation 
of this property. Furthermore, it allows us to predict and explain the occurrence of ``flipped'' profiles of the mean phase velocities.

To this purpose, we study the initial dynamics that is simplified by the
initial conditions. The amplitude initial conditions, Eq.\,(\ref{eq:9}),
result in vanishing local dynamics of the amplitudes, $F(r,t_{0})=0$,
and the initial dynamics is simplified to 

\begin{eqnarray}
\partial_{t}r(x,t_{0}) & = & \Sigma_{r}(x,t_{0}),\label{eq:10}\\
\partial_{t}\theta(x,t_{0}) & = & \Sigma_{\theta}(x,t_{0}).\label{eq:11}
\end{eqnarray}
Using the initial conditions, Eqs.\,(\ref{eq:9}) and (\ref{eq:91}),
in the definitions of the coupling terms given by Eqs.\,(\ref{eq:5}) and (\ref{eq:6}), the initial coupling terms 
are simplified to

\begin{eqnarray}
\Sigma_{r}(x,t_{0}) & = & -\sigma\cos(\alpha)C_{r}(x),\label{eq:12}\\
\Sigma_{\theta}(x,t_{0}) & = & -\sigma\sin(\alpha)C_{\theta}(x),\label{eq:13}
\end{eqnarray}
where the function $C_{r}(x)$ summarizes the values of the integral
in the amplitude dynamics and other constants, and the function $C_{\theta}(x)$
summarizes the values of the integral in the phase dynamics and other
constants. 
If we now take a look at the scenario sketched in Fig.\,\ref{fig:1}, the mechanism leading to a chimera state is uncovered: While the functions representing the integral vanish towards the center of the synchronized domains, leading to synchronized behavior, their non-zero values towards the borders between the anti-phase domains leads to varying, desynchronized behavior.

For $\alpha$ close to $\pi /2$ the amplitude coupling term $\Sigma_{r}(x,t_0)$ nearly vanishes 
and the magnitude of the phase coupling term $\Sigma_{\theta}(x,t_0)$ is maximum, thus effectively restricting the variation to the phases. It is important to note that this effect of the initial coupling terms in Eqs.\,(\ref{eq:12}) and (\ref{eq:13}) also occurs if the coupling phase $\alpha$ approaches the value $-\pi/2$. This property is used in the next subsection where the occurrence of "flipped" profiles of the mean phase velocities and its connection to the coupling phase $\alpha$ is discussed. The possibility to increase amplitude modulations by a proper choice of coupling strength $\sigma$ is analyzed in the subsequent section.

\subsection{``Flipping'' profiles of the mean phase velocities}

\begin{figure}
\begin{centering}
\includegraphics[width=0.5\textwidth]{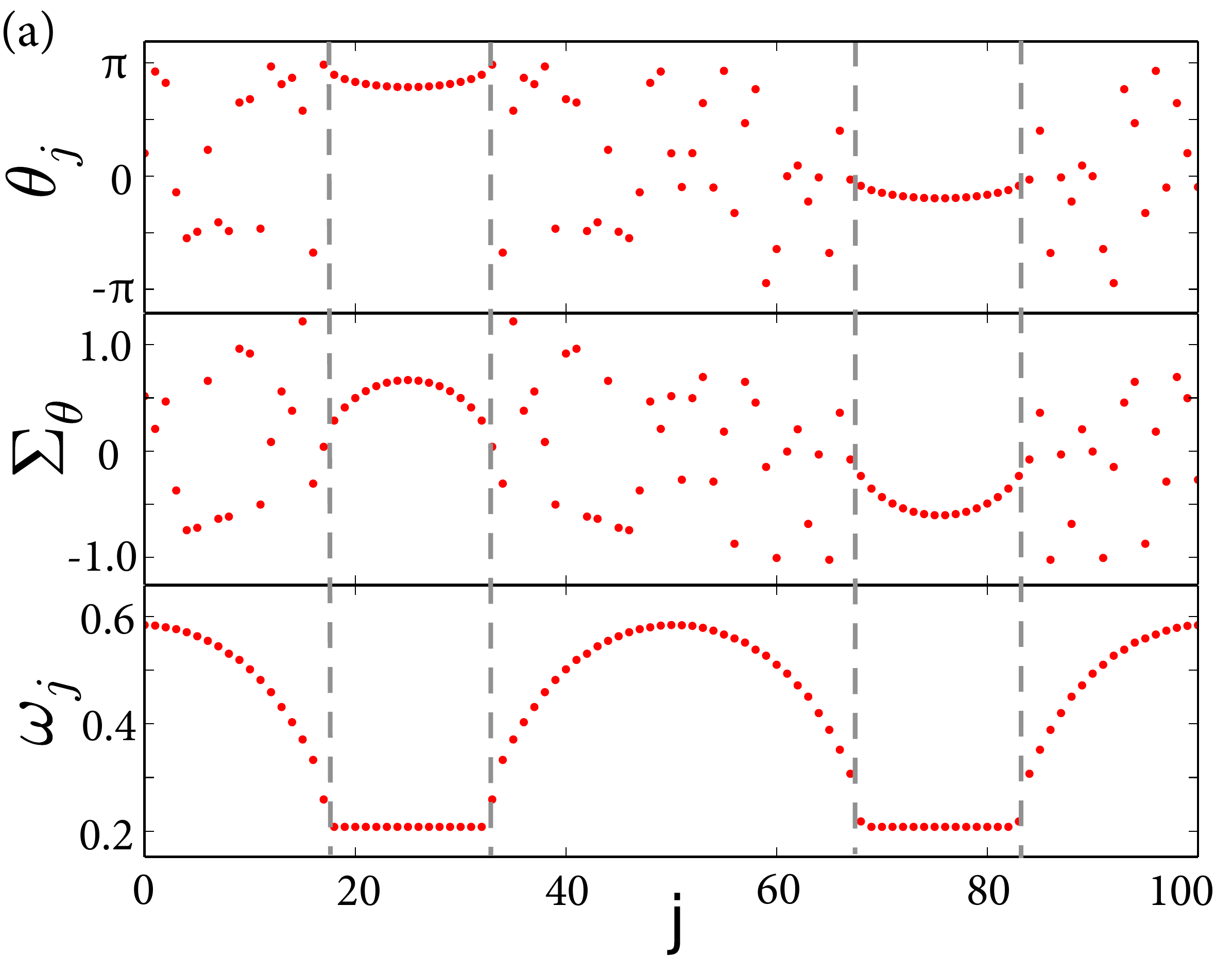}
\includegraphics[width=0.5\textwidth]{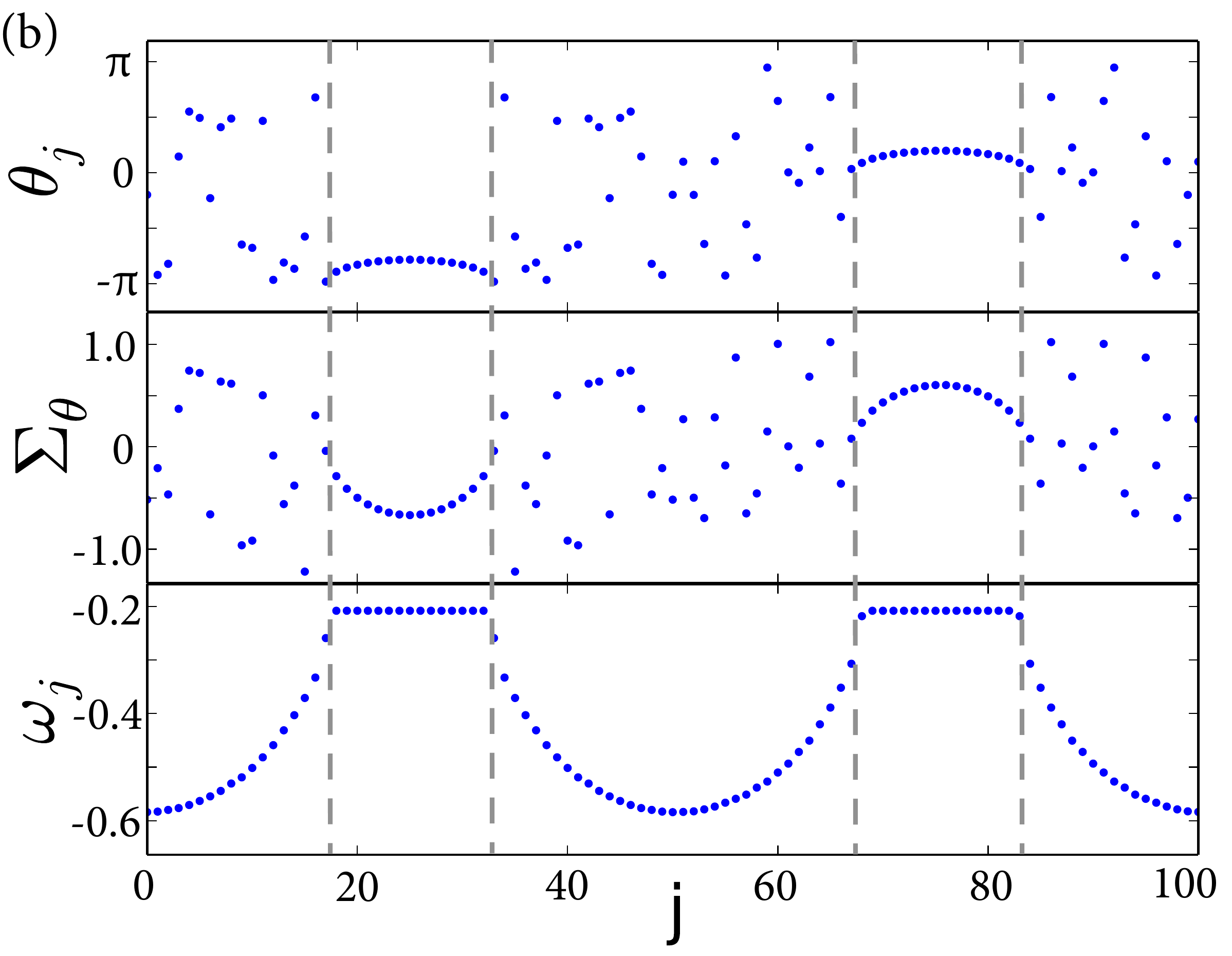}
\par\end{centering}

\centering{}\caption{Snapshots of the phases $\theta_j$ (top panels), phase coupling term $\Sigma_{\theta}$ (middle panels)
and profile of the mean phase velocities $\omega_j$ (bottom panels) 
at $t=400$ for (a) $\alpha=-(\pi/2-0.15)$ and (b) $\alpha=\pi/2-0.15$.
Initial conditions and parameters as in Fig.\,\ref{fig:1-1-3}. \label{fig:1-1}}
\end{figure}

From Eq.\,(\ref{eq:11}) it follows that the sign of the
phases is determined by the phase coupling term $\Sigma_{\theta}(x,t)$
solely. Therefore, a change in the sign of $\Sigma_{\theta}$ changes the phase dynamics qualitatively. In particular, for positive values 
of $\Sigma_{\theta}$ the phases are expected
to evolve to positive values while for negative values of $\Sigma_{\theta}$
the phases become negative. In the first case, a positive
phase velocity results in a normal concave ``upside'' profile of the mean
phase velocities $\omega_j=\partial_{t}\theta(x_j)$, while in the latter case negative values of the phase
velocities lead to a convex ``flipped'' profile of the
mean phase velocities, see Fig.\,\ref{fig:1-1}.

The sign of $\Sigma_\theta$ is changed by a suitable choice of $\alpha$. Coupling phases $\alpha$ 
close to $-\pi/2$ fulfill the requirement of almost vanishing amplitude coupling terms $\Sigma_r$ and maximum magnitude of the phase coupling terms $\Sigma_\theta$, as well. Taking advantage of this, the sign of the coupling terms can be modified by a change of the sign of $\alpha$. As shown in Fig.\,\ref{fig:1-1}a a value of $\alpha=\pi/2-0.15$ leads to a negative sign of the coupling terms $\Sigma_\theta$, and a "flipped" profile of the mean phase velocities can be observed for the domains of incoherent phases. In contrast, in Fig.\,\ref{fig:1-1}b a choice of $\alpha=-(\pi/2-0.15)$ results in a positive coupling term $\Sigma_\theta$, leading to a normal concave "upside" profile of positive mean phase velocities.

\subsection{Transition from phase to amplitude-phase chimera states\label{sub:Impact-of-coupling}}

\begin{figure}
\begin{centering}
\includegraphics[width=0.51\textwidth]{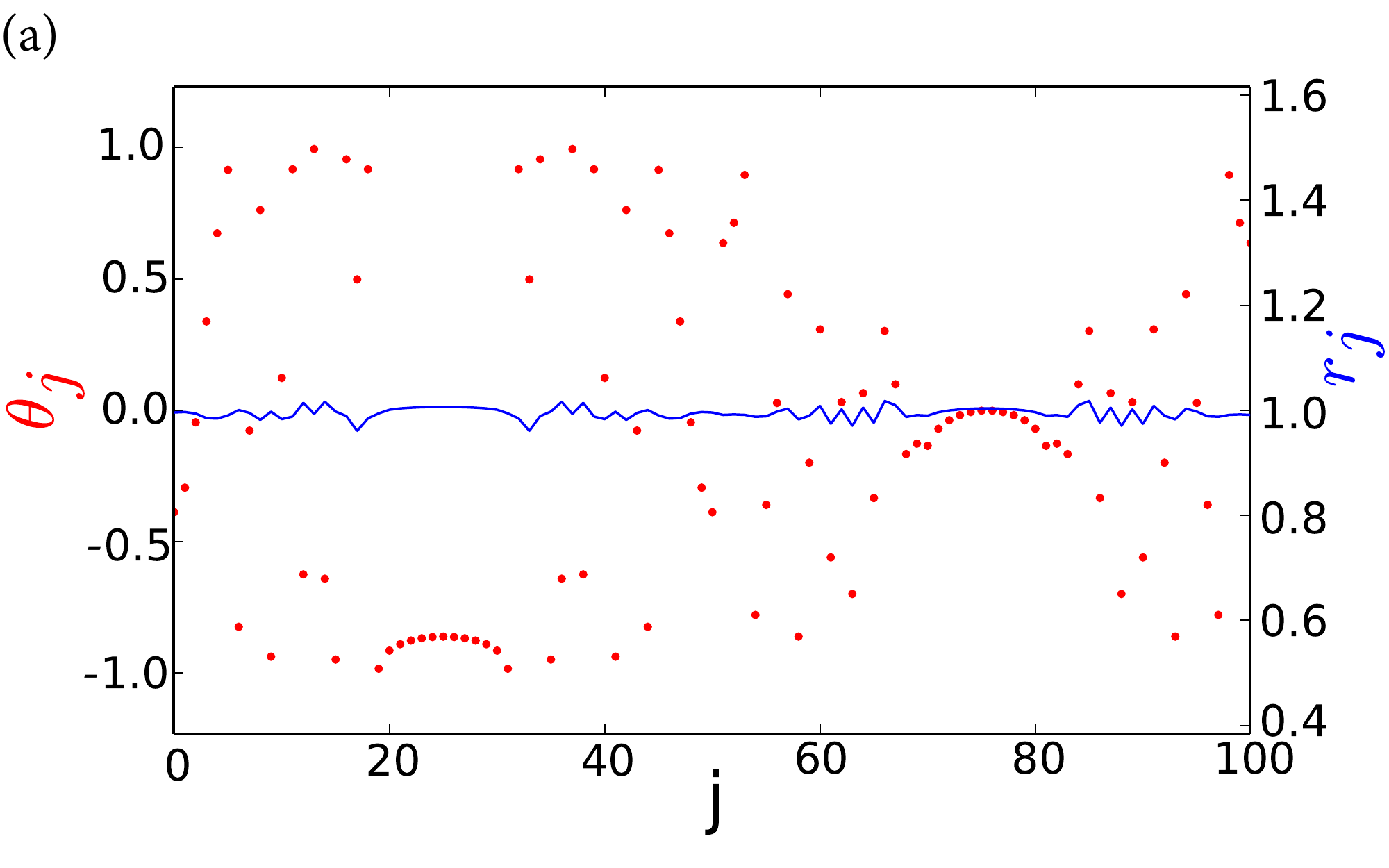}
\includegraphics[width=0.51\textwidth]{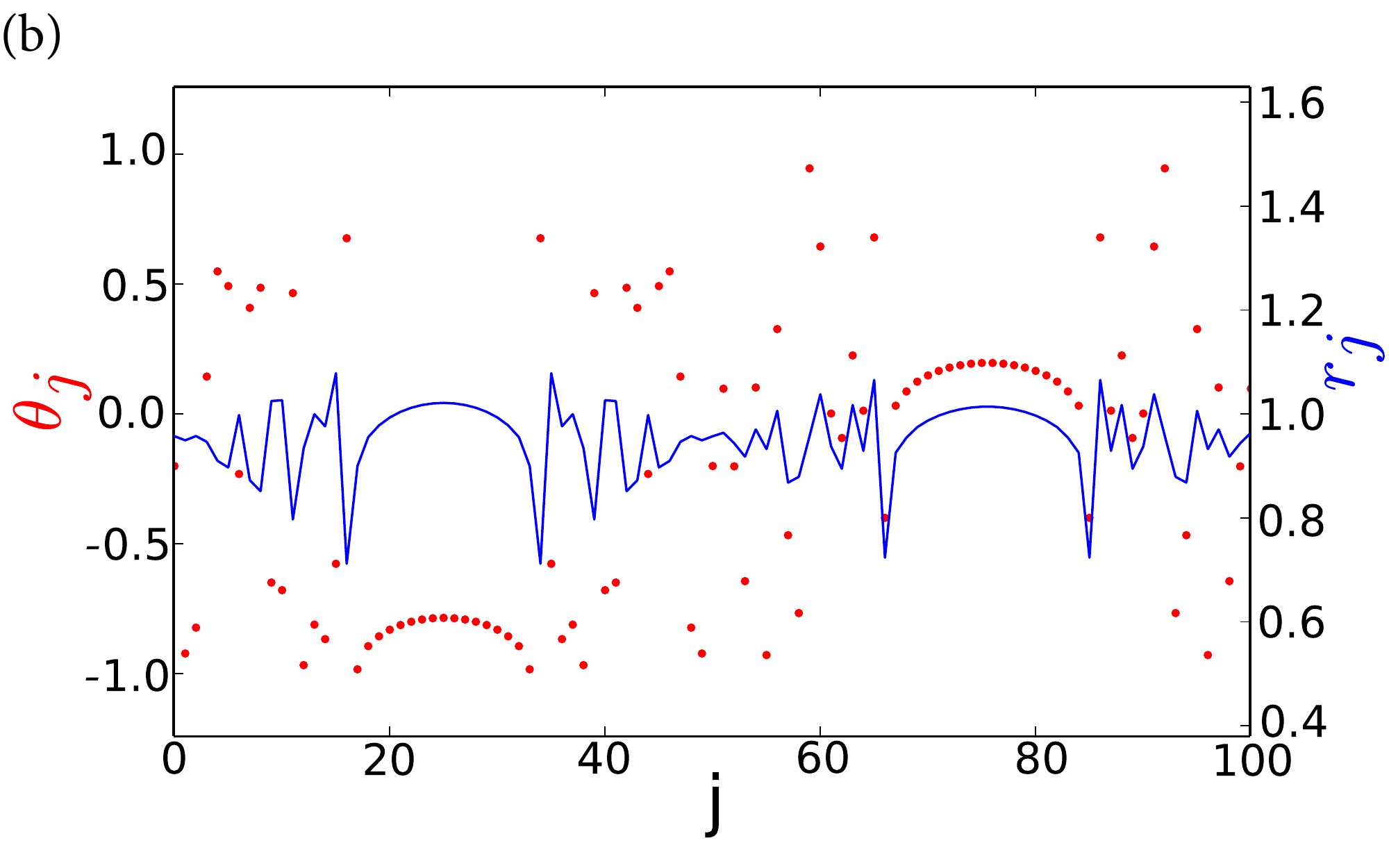}
\par
\end{centering}

\caption{Snapshots of the phase $\theta_{j}$ (red dots) and amplitudes $r_{j}$ (blue lines) at $t=2000$ for (a) weak coupling strength, $\sigma=0.1$, and (b) increased coupling strength, $\sigma=0.6$.
Initial conditions and other parameters as in Fig.\,\ref{fig:1-1-3}.\label{fig:1-2}}
\end{figure}

A feature of amplitude-mediated chimera states,
as reported recently\cite{SET13}, is the coexistence of coherent and incoherent domains not only for the phases but also for the amplitudes. By inspecting the simplified
coupling term in the amplitudes, $\Sigma_{r}$, it is possible to
explain the transition from phase chimera states to amplitude-phase
chimera states by increasing the coupling strength $\sigma$.
As discussed above, the initial dynamics for the amplitudes is simplified
to

\begin{equation}
\partial_{t}r(x,t_0)=\Sigma_{r}(x,t_0),
\end{equation}
where the coupling term for the amplitudes is given by 
\begin{equation}
\Sigma_{r}(x,t_0)=-\sigma\cos(\alpha)C_{r}(x).
\end{equation}
The magnitude of the coupling term $\Sigma_{r}$ increases linearly by
the coupling strength $\sigma$. Therefore, in the limit of weak coupling
($\sigma=0.1$) the occurrence of a phase chimera is expected, where
the variations in the amplitudes are negligible, see Fig.\,\ref{fig:1-2}a.
In contrast, as shown in Fig.\,\ref{fig:1-2}b, for increased values of the coupling strength ($\sigma=0.6$)
the amplitude variations increase and the incoherent dynamics of the phases is combined with non-vanishing modulations in the amplitudes $r_j$.

\section{Conclusion}

In the current study, we have analyzed chimera states in networks of Stuart-Landau oscillators. We have provided an analytical argument that explains the need for an off-diagonal coupling, i.e., a phase-lag in the coupling, in order to create chimera states. Based on this, we have discussed the impact of the sign of the coupling phases. We were able to show how the sign of the coupling phase determines the sign of the profile of the mean phase velocities. Furthermore, we exemplified how our argument gives an intuitive explanation for the transition from phase chimera states in the limit of weak coupling to a state sharing the main features of an amplitude-mediated chimera state in the case of intermediate coupling strength.

\begin{acknowledgments}
 This work was supported by Deutsche Forschungsgemeinschaft in the framework of Collaborative Research Center SFB 910. 
\end{acknowledgments}



\end{document}